\title{Reputation for Playing Mixed Actions: A Characterization Theorem}
\author{Harry PEI\footnote{Department of Economics, Northwestern University. This article is based on Chapter 1 of my dissertation at MIT,
which is split into the current paper (Chapter 1.3) and
``Reputation Effects under Interdependent Values'' (Chapter 1.4). I am indebted to Daron Acemoglu, Drew Fudenberg, Juuso Toikka, and Alex Wolitzky for guidance and support. I thank Heski Bar-Isaac, Daniel Clark, Martin Cripps, Joyee Deb, Mehmet Ekmekci, Jack Fanning,
Yuhta Ishii, Elliot Lipnowski, Qingmin Liu, Shuo Liu, Lucas Maestri, Marcin P\c{e}ski,
Bruno Strulovici, Can Urgun, Nicolas Vieille, Geyu Yang, an associate editor, and two anonymous referees for helpful comments.}}
\date{April 3, 2021}
\documentclass[11pt]{article}

\usepackage{amsmath}
\usepackage{graphicx}
\usepackage{amsfonts}
\usepackage{amsthm}
\usepackage{setspace}
\usepackage{tikz}
\usepackage{amssymb}
\usetikzlibrary{patterns}
\usepackage{sgame}
\usepackage{color}
\usepackage{hyperref}
\usepackage{times}
\usepackage{enumitem}
\usepackage{csquotes}
\usepackage{multirow,array}
\usepackage[top=1in, bottom=1in, left=0.9in, right=0.9in]{geometry}
\begin{document}
\numberwithin{equation}{section}

\maketitle
\noindent \textbf{Abstract:} A patient player privately observes a persistent state that directly affects his myopic opponents' payoffs, and can be one of the several commitment types that plays the same mixed action in every period. I characterize the set of environments under which the patient player obtains at least his commitment payoff in all equilibria regardless of his stage-game payoff function. Due to interdependent values, the patient player cannot guarantee his mixed commitment payoff by imitating the mixed-strategy commitment type, and small perturbations to a pure commitment action can significantly reduce the patient player's guaranteed equilibrium payoff.\\


\newtheorem{Proposition}{\hskip\parindent\bf{Proposition}}
\newtheorem{Theorem}{\hskip\parindent\bf{Theorem}}
\newtheorem{Lemma}{\hskip\parindent\bf{Lemma}}[section]
\newtheorem{Corollary}{\hskip\parindent\bf{Corollary}}
\newtheorem{Definition}{\hskip\parindent\bf{Definition}}
\newtheorem{Assumption}{\hskip\parindent\bf{Assumption}}
\newtheorem{Condition}{\hskip\parindent\bf{Condition}}
\newtheorem{Claim}{\hskip\parindent\bf{Claim}}
\newtheorem*{Assumption1}{\hskip\parindent\bf{Assumption 1'}}

\begin{spacing}{1.5}
\section{Introduction}\label{sec1}
I examine patient players' returns from building reputations for playing mixed actions. To fix ideas, consider a profit-maximizing firm that needs to decide whether
to imitate the behavior of an ethical firm that intrinsically cares about its worker and customers.
Suppose the ethical firm commits to provide good customer service unless its worker is sick, and consumers can only observe the quality of service but not whether the worker is sick or healthy, then the ethical firm  behaves \textit{as if} it is mixing between providing good service and bad service.

The reputation results in Fudenberg and Levine (1989, 1992)
imply that when consumers' payoffs depend only on their actions and the firm's action, there is no qualitative difference between establishing reputations for playing pure actions and that for playing mixed actions.
By playing the commitment action in every period, a patient firm receives at least its commitment payoff from that action, and
under generic parameter values, perturbing a pure commitment action leads to a continuous change in the firm's lowest equilibrium payoff.

This paper shows that whether the commitment action is pure or mixed has significant effects on a patient player's payoff in interdependent value environments.
I study a repeated game between a patient player $1$ (e.g., firm) and an infinite sequence of myopic player $2$s (e.g., consumers).
Player $1$ privately observes the realization of a payoff-relevant state (e.g., product safety or durability) that is constant over time and affects both players' stage-game payoffs, in addition to knowing whether he is rational or committed. The rational player $1$
maximizes his discounted average payoff. The committed player $1$ mechanically plays the same \textit{commitment
action} in every period, which can be pure or mixed and can depend on the persistent state. This differs from Pei (2020) which assumes that all commitment types play pure strategies.
Player $2$s can observe all the actions taken in the past but cannot observe the state or player $1$'s mixed actions.

My result
 characterizes the set of interdependent value environments under which the patient player
receives at least his commitment payoff in every equilibrium
regardless of his stage-game payoff function. My characterization implies that securing commitment payoffs from mixed actions requires more demanding conditions than securing commitment payoffs from pure actions, and that
small perturbations to a pure commitment action can
significantly reduce the patient player's lowest equilibrium payoff.

Intuitively, when a commitment action is mixed, some pure actions in the support of this mixed commitment action can be played with strictly higher probability by some strategic types than by the mixed-strategy commitment type. If that is the case, then playing these pure actions \textit{increases} the likelihood ratios between these strategic types and the corresponding commitment type.
This stands in contrast to the case where the commitment action is pure, under which the likelihood ratio between every strategic type and the pure-strategy commitment type cannot increase when the patient player plays that pure commitment action.

My analysis unveils another difference between private and interdependent values, that when player $2$'s best reply to the commitment action depends on the state, player $1$ \textit{cannot} secure his mixed commitment payoff by imitating the mixed-strategy commitment type.
To the best of my knowledge, this observation is novel in the reputation literature, since all existing reputation results  are shown by bounding a patient player's payoff when he imitates some commitment type. Intuitively,
playing some actions in the support of a mixed commitment strategy can increase the likelihood ratio between some strategic types and this commitment type, and furthermore, it can trigger negative inferences about the state.
Since the state is persistent and affects player $2$'s best reply to the commitment action, player $2$'s belief about the state in any given period can have a long-lasting effect on player $1$'s continuation value. This happens when
all strategic types play the commitment action after that period, in which case learning stops on the equilibrium path and
player $2$'s belief  about the state in that period determines her best reply in all subsequent periods.
This suggests the need for player $1$ to
\textit{take actions selectively} in the support of the mixed commitment action in order to avoid such negative inferences.


However, taking actions selectively instead of playing the mixed commitment action raises two new concerns. First,  player 1 may  play some low-payoff actions too frequently,
in which case his expected payoff may fall below
 his commitment payoff.
Second, given that player 1 may not play the mixed commitment action in every period,
he may fail to convince his opponents that the commitment action will be played in the future.

I establish a learning result that addresses both concerns. It shows that for every strategy profile, player 1 can find a deviation under which
(1) player $2$ has an incentive to play the desirable best reply to the commitment action in every period under her posterior belief about the state;
(2) with probability close to one, the discounted frequency of player $1$'s action is close to the mixed commitment action;
(3) in expectation, player $2$ believes that player $1$'s action is close to the mixed commitment action in all except for a bounded number of periods. My proof uses a combination of the Doob's upcrossing inequality, the central limit theorem for triangular sequences, and the entropy techniques in Gossner (2011). My approach can also be used to derive lower bounds on the patient player's equilibrium payoff in reputation games with imperfect monitoring.

This paper contributes to the reputation literature by examining players' guaranteed returns from building reputations
when values are interdependent and their opponents \textit{cannot} perfectly monitor whether they have honored their commitment. It highlights the differences between building reputations for playing mixed actions and that for playing pure actions, and explain why these differences are caused by interdependent values.



My analysis unveils the challenges to build reputations when learning is \textit{confounded}. Even though the informed player can convince his opponents about his future actions, he may not teach them how to best reply when their payoff functions depend on a persistent state.
This is related to the recent works of Yang (2019) and Deb and Ishii (2021), in which confounded learning is caused by uncertainty in the monitoring structure.

Yang (2019) focuses on private value environments and provides sufficient conditions under which the patient player can secure his commitment payoff.
Deb and Ishii (2021) allow for
interdependent values and
uncertainty in the monitoring structure. They assume that
for every pair of states, there exists an action of the long-run player
such that the distribution over public signals induced by this action in the first state is different from that induced by any action in the second state. Their identification condition is violated in my model where the uninformed players learn about the informed player's type \textit{only} through the latter's actions.


Ekmekci and Maestri (2019) and Ekmekci, Gorno, Maestri, Sun and Wei (2021) obtain sharp predictions on an informed player's payoff when monitoring is imperfect and a long-lived uninformed player decides whether to continue to interact with the informed player or to irreversibly stop the interaction.
By contrast, my result
highlights the challenges to build reputations when the uninformed players can freely choose their actions.


\section{Model}\label{sec2}
Time is discrete, indexed by $t=0,1,2...$. A long-lived player $1$ (he, e.g., a seller) with discount factor $\delta \in (0,1)$ interacts with an infinite sequence of short-lived player $2$s (she, e.g., consumer), arriving one in each period and each plays the game only once. In period $t$, players simultaneously choose their actions $(a_{1,t},a_{2,t}) \in A_1 \times A_2$.

Player $1$ has private information about (1) a payoff-relevant state $\theta \in \Theta$, and (2) whether he is \textit{strategic} or \textit{committed}. Both are drawn and fixed before period $0$.
If player $1$ is strategic, then he can flexibly choose his actions in order to maximize his discounted average payoff.
If player $1$ is committed, then he mechanically follows
one of the several \textit{commitment plans}. A typical commitment plan is denoted by
$\gamma: \Theta \rightarrow \Delta (A_1)$, according to which
the committed player
plays $\gamma(\theta) \in \Delta (A_1)$ in every period when the realized state is $\theta$.
Let $\Gamma$ be an exogenous set of \textit{feasible commitment plans} that the committed player $1$ can follow. Let
\begin{equation}\label{2.1}
  \mathcal{A}_1^* \equiv \{ \alpha_1 \in  \Delta(A_1) | \textrm{ there exist } \gamma \in \Gamma \textrm{ and } \theta \in \Theta \textrm{ such that } \gamma (\theta)=\alpha_1 \} \subset \Delta (A_1),
\end{equation}
be the set of \textit{commitment actions}.
Intuitively,  $\alpha_1^*$ belongs to $\mathcal{A}_1^*$ if and only if $\alpha_1^*$ is played in some state under some feasible commitment plan.
Let $\gamma^*$ stand for player $1$ being strategic.
Let
\begin{equation}\label{2.2}
    \mu \in \Delta \Big(\Theta \times   \underbrace{\big(\{\gamma^*\} \cup \Gamma \big)}_{\textrm{player 1's characteristics}}   \Big)
\end{equation}
be player $2$'s prior belief, which is a joint distribution of the state $\theta$ and player $1$'s \textit{characteristics}, namely, whether he is strategic or committed, and if he is committed, which feasible plan in $\Gamma$ he follows.

For every $\theta \in \Theta$,
I say that player $1$ is \textit{strategic type $\theta$} if he is strategic and knows that the state is $\theta$.
For every $\alpha_1^* \in \mathcal{A}_1^*$, I say that player $1$ is \textit{commitment type $\alpha_1^*$} if he is committed and plays $\alpha_1^*$ in every period.
Let $\mu(\theta)$ be the prior probability of \textit{strategic type} $\theta$, and
let $\mu(\alpha_1^*)$ be the prior probability of \textit{commitment type} $\alpha_1^*$.
\begin{Assumption}\label{Ass1}
Sets $\Theta$, $\Gamma$, $A_1$, and $A_2$ are finite, $|A_1|, |A_2|\geq 2$, and $\mu$ has full support.
\end{Assumption}
Cases in which $|A_1|=1$ or $|A_2|=1$ are trivial since either player $1$ or player $2$
has no choice to make. Throughout the paper, I use $m \equiv |\Theta|$ to denote the number of states.

Let $h^t \equiv \{a_{1,s},a_{2,s}\}_{s =0}^{t-1} \in \mathcal{H}^t$ be a public history.
Let $\mathcal{H} \equiv \bigcup_{t=0}^{+\infty} \mathcal{H}^t$ be the set of public histories.
Player $1$'s private history consists of the public history and his persistent private information.
Player $2$'s private history coincides with the public history.
Let $\sigma_1 \equiv (\sigma_{\theta})_{\theta \in \Theta}$ be strategic player $1$'s strategy, with
$\sigma_{\theta}: \mathcal{H} \rightarrow \Delta(A_1)$.
Let $\sigma_2: \mathcal{H} \rightarrow \Delta(A_2)$ be player $2$'s strategy. Let $\sigma \equiv\big( \sigma_1, \sigma_2 \big)$ be a strategy profile, with $\sigma \in \Sigma$.


For $i \in \{1,2\}$, player $i$'s stage-game payoff in period $t$ is $u_i(\theta,a_{1,t},a_{2,t})$, which is naturally extended to mixed actions.
This formulation allows for interdependent values since $u_2$ depends on $\theta$, which is player $1$'s private information.
For every $\phi \in \Delta(\Theta)$, $\alpha_1 \in \Delta (A_1)$, and $u_2: \Theta \times A_1 \times A_2 \rightarrow \mathbb{R}$, let
\begin{equation}\label{2.3}
    \textrm{BR}_2(\phi,\alpha_1|u_2) \equiv \arg \max_{a_2 \in A_2} \Big\{
    \sum_{\theta \in \Theta} \sum_{a_1 \in A_1} \phi(\theta) \alpha_1(a_1) u_2(\theta,a_1,a_2)
    \Big\},
\end{equation}
be the set of pure best replies to $\alpha_1$ when $\theta$ is distributed according to $\phi$.
Let $\textrm{BR}_2(\theta,\alpha_1|u_2) \subset A_2$ be player $2$'s pure best replies to  $\alpha_1$ when the state is $\theta$. I make the following assumption that is satisfied for generic $u_2$:
\begin{Assumption}\label{Ass2}
For every $\alpha_1^* \in \mathcal{A}_1^*$ and $\theta \in \Theta$,
$\textrm{BR}_2(\theta, \alpha_1^*| u_2)$ is a singleton.
\end{Assumption}
Let $a_2^* (\theta,\alpha_1^*|u_2)$ be player $2$'s unique best reply to $\alpha_1^*$ in state $\theta$ when her payoff function is $u_2$.


\section{Characterization Theorem}\label{sec3}
I provide sufficient and (almost) necessary conditions under which the patient player can secure his commitment payoff \textit{regardless of} his stage-game payoff function.\footnote{My conditions are \textit{almost} necessary since (1) they leave out a knife-edge set of type distributions, (2) they require an additional generic assumption on players' stage-game payoffs, and (3) in the case where $\alpha_1^*$ is mixed, they also impose additional restrictions on the convex hull of player $1$'s commitment actions.} My analysis highlights the distinctions between
building reputations for playing mixed actions and building reputations for playing pure actions when values are interdependent, both in terms of the patient player's guaranteed equilibrium payoff and the behavior under which he is guaranteed to receive his commitment payoff.

\subsection{Commitment Payoff \& Private Value Benchmark}
For every $(\theta,\alpha_1^*) \in \Theta \times \mathcal{A}_1^*$,
player $1$'s (complete information) \textit{commitment payoff} from playing $\alpha_1^*$ in state $\theta$ is:
\begin{equation}\label{3.1}
   v_{\theta} (\alpha_1^* , u_1,u_2) \equiv  u_1 \Big(\theta,\alpha_1^*,a_2^* (\theta,\alpha_1^*|u_2) \Big).
\end{equation}
Let $\textrm{NE}(\delta,\mu,u_1,u_2) \subset \Sigma$
be the set of Bayes Nash equilibria.
Let
\begin{equation}\label{3.2}
\underline{v}_{\theta}(\delta,\mu , u_1,u_2) \equiv \inf_{\sigma \in \textrm{NE}(\delta,\mu,u_1,u_2)} \mathbb{E}^{(\sigma_{\theta},\sigma_2)} \Big[ \sum_{t=0}^{+\infty} (1-\delta)\delta^t u_1(\theta,a_{1,t},a_{2,t})\Big]
\end{equation}
be strategic type $\theta$'s \textit{lowest equilibrium payoff}. I examine for given $(\theta,\alpha_1^*) \in \Theta \times \mathcal{A}_1^*$, when it is the case that
\begin{equation}\label{3.3}
  \underbrace{  \liminf_{\delta \rightarrow 1}  \underline{v}_{\theta}(\delta,\mu,u_1,u_2)}_{\textrm{strategic type $\theta$'s lowest equilibrium payoff}} \geq \underbrace{v_{\theta}(\alpha_1^*,u_1,u_2)}_{\textrm{strategic type $\theta$'s commitment payoff from $\alpha_1^*$}} \textrm{ for all } u_1?
\end{equation}

In private value environments where $u_2$ \textit{does not} depend on the state, or more generally,
player $2$'s best reply to $\alpha_1^*$ does not depend on the state, inequality (\ref{3.3}) is implied by the results in Fudenberg and Levine (1989, 1992) and player $1$ can guarantee his commitment payoff by playing $\alpha_1^*$ in every period. The intuition is that
after observing player $1$'s action frequency matches $\alpha_1^*$ for a long time, player $2$s will be convinced that player $1$'s
action will be close to $\alpha_1^*$
in all future periods, so they will play a myopic best reply to $\alpha_1^*$. As a result, the patient player receives at least his commitment payoff
in all except for a bounded number of periods.
Since
$\textrm{BR}_2(\theta,\alpha_1^*| u_2)$
is a singleton, player $1$'s commitment payoff is continuous at $\alpha_1^*$, which means that
a small perturbation to a pure commitment action leads to a continuous change in player 1's lowest equilibrium payoff.

\subsection{A Motivating Example}\label{sub3.2}
I use an example to show that the aforementioned conclusions under private values are no longer true when player $2$'s best reply to $\alpha_1^*$ depends on the state. Suppose $\Theta \equiv \{\theta^*,\theta_1,\theta_2\}$ and players' stage-game payoffs are
\begin{center}
\begin{tabular}{| c | c | c | c |}
  \hline
  $\theta^*$ & $G$ & $M_1$ & $M_2$\\
    \hline
    $H$ & $1,3$ & $-\frac{1}{2},0$ & $-\frac{1}{2},0$\\
  \hline
  $I$ & $2,-1$ & $0,-\frac{1}{2}$ & $0,-\frac{1}{2}$ \\
  \hline
  $L$ & $3,-\frac{3}{2}$ & $\frac{1}{2},-1$ & $\frac{1}{2},-1$ \\
  \hline
\end{tabular}
\begin{tabular}{| c | c | c | c |}
  \hline
  $\theta_1$ & $G$ & $M_1$ & $M_2$\\
    \hline
    $H$ & $2,\frac{1}{2}$ & $1,\frac{3}{2}$ & $-1,0$\\
  \hline
  $I$ & $2,0$ & $1,1$ & $-1,-\frac{1}{2}$ \\
  \hline
  $L$ & $3,-1$ & $\frac{3}{2},-1$ & $0,-1$ \\
  \hline
\end{tabular}
\begin{tabular}{| c | c | c | c |}
  \hline
  $\theta_2$ & $G$ & $M_1$ & $M_2$\\
    \hline
    $H$ & $2,\frac{1}{2}$ &   $-1,0$ & $1,\frac{3}{2}$\\
  \hline
  $I$ & $2,0$ &  $-1,-\frac{1}{2}$ & $1,1$ \\
  \hline
  $L$ & $3,-1$  & $0,-1$ & $\frac{3}{2},-1$ \\
  \hline
\end{tabular}
\end{center}

For an economic interpretation of this game, suppose player $1$ is a firm that chooses between high ($H$), intermediate ($I$), and low ($L$) effort that determines the quality of its customer service, and player $2$ is a consumer that chooses between a good product ($G$), a mediocre product with the first characteristic ($M_1$), and a mediocre product with the second characteristic ($M_2$). If the state is $\theta^*$, then
exerting high effort is costly for the firm and
purchasing the good product is worthwhile
for the consumers if and only if the firm exerts high effort.
 If the state is $\theta_i$ with $i \in \{1,2\}$, then it is not worthwhile for the consumer to buy the good product and exerting high effort is not costly for the firm compared to exerting intermediate effort. However, the characteristics of the two mediocre products
 are important for the consumers in those states. In state $\theta_i$,
each consumer strictly prefers the mediocre product with characteristic $i$ as long as the firm does not exert low effort.

First, consider a benchmark scenario in which all commitment actions are pure. Suppose (1) $\mathcal{A}_1^*=\{H,L\}$, (2) conditional on player $1$ being commitment type $H$, state $\theta^*$ occurs with probability $1$, and (3) player $2$'s prior belief $\mu$ is such that $\mu(\theta_1)=\mu(\theta_2)=3 \mu (H)$. According to Theorem 1' of Pei (2020, page 2191), type $\theta^*$  receives at least his commitment payoff from $H$ (equals $1$) in every equilibrium when he is patient.

Next, consider a perturbed game in which $\mathcal{A}_1^* \equiv \{\alpha_1^*,L\}$, where $\alpha_1^*=(1-\varepsilon) H +\varepsilon I$ and $\varepsilon>0$ is sufficiently small.
Suppose the type distribution $\mu$ is such that
$\mu(\theta_1)=\mu(\theta_2)=3 \mu (\alpha_1^*)$, and
conditional on player $1$ being commitment type
$\alpha_1^*$, state $\theta^*$ occurs with probability $1$.
I make no restriction on the probability of strategic type $\theta^*$ except that it cannot occur with probability $1$, i.e., it can occur with probability arbitrarily close to $1$.
Consider the following equilibrium in which type $\theta^*$'s payoff is $1/2$ no matter how patient he is.
\begin{itemize}
\item Strategic type $\theta^*$ plays $L$ in every period. In period $0$, strategic type $\theta_1$ plays $H$ and strategic type $\theta_2$ plays $I$. Starting from period $1$, strategic types $\theta_1$ and $\theta_2$ play $(1-\varepsilon) H +\varepsilon I$.
\item In period $0$, player $2$ plays $M_1$ if $\mu(\theta_1) \geq \mu(\theta_2)$ and plays $M_2$ otherwise.
If she observes $L$ in period $0$, then she plays $\frac{1}{2}M_1+ \frac{1}{2}M_2$ starting from period $1$.
If she observes $H$ in period $0$, then she plays $M_1$ starting from period $1$ on the equilibrium path. If she observes $I$ in period $0$, then she plays $M_2$ starting from period $1$ on the equilibrium path. If player $1$ plays $L$ after $H$ or $I$, then player $2$'s posterior belief assigns probability $1/2$ to strategic type
$\theta_1$, and probability $1/2$ to strategic type $\theta_2$, after which she plays $\frac{1}{2}M_1+ \frac{1}{2}M_2$ in every subsequent period.
\end{itemize}

The existence of low-payoff equilibrium stands in contrast to
the conclusions in the private value benchmark and in interdependent value games where all commitment types play pure strategies.
This is because in state $\theta^*$, player $1$'s commitment payoff from $\alpha_1^*$ is strictly greater than that from $H$. In a private value reputation game where state $\theta^*$ occurs with probability $1$,
the patient player's guaranteed equilibrium payoff increases once we replace commitment type $H$ with commitment type $\alpha_1^*$.
However, when state $\theta^*$ occurs with probability close to $1$, replacing commitment type $H$ with commitment type $\alpha_1^*$ while keeping other parameters unchanged leads to a significant decrease in player $1$'s lowest equilibrium payoff.

This low-payoff equilibrium is driven by the presence of interdependent values and mixed-strategy commitment types. First, player $2$'s best reply to $\alpha_1^*$ is $G$ if and only if  $\theta=\theta^*$. Therefore, convincing her that action $\alpha_1^*$ will be played in the future does not determine her best reply. This effect is also present in Pei (2020).

Second, when the commitment type plays a nontrivially mixed action, it can be the case that playing every action in the support of this mixed action leads to a negative inference about the payoff-relevant state.
This stands in contrast to Pei (2020) since
when the commitment action is pure, the likelihood ratio between every strategic type and that commitment type can never increase when player $1$ plays this pure commitment action.

In the constructed equilibrium, $G$ is player $2$'s best reply to $\alpha_1^*$ under her prior belief about the state, from which player $1$ receives a high payoff. If player $1$ plays $H$ in period $0$, then the likelihood ratio between strategic type $\theta_1$ and commitment type $\alpha_1^*$ increases; if player $1$ plays $I$ in period $0$, then the likelihood ratio between strategic type $\theta_2$ and commitment type $\alpha_1^*$ increases; if player $1$ plays $L$ in period $0$, then he is separated from commitment type $\alpha_1^*$.
I depict player $2$'s belief in Figure 1.
In summary, no matter which action player $1$ takes in period $0$, $G$ is no longer player $2$ best reply to $H$ under her posterior belief about the state.


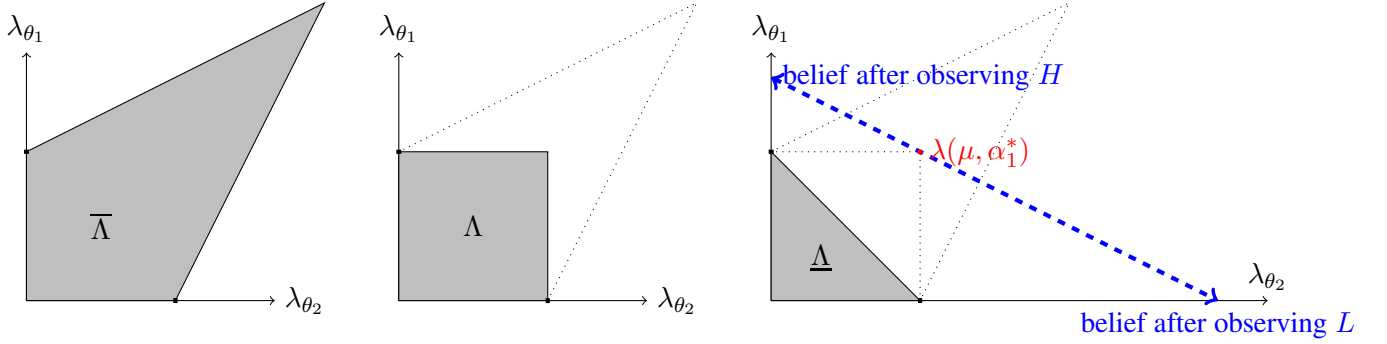
\begin{figure}\label{Figure1}
\begin{center}
\begin{tikzpicture}[scale=0.33]
\draw [<->] (-15,10)node[above]{$\lambda_{\theta_1}$}--(-15,0)--(-5,0)node[right]{$\lambda_{\theta_2}$};
\draw [fill=lightgray] (-15,0)--(-9,0)--(-3,12)--(-15,6)--(-15,0);
\node at (-12,3) {$\overline{\Lambda}$};
\draw [ultra thick] (-9,0.1)--(-9,-0.1);
\draw [ultra thick] (-14.9,6)--(-15.1,6);
\draw [<->] (0,10)node[above]{$\lambda_{\theta_1}$}--(0,0)--(10,0)node[right]{$\lambda_{\theta_2}$};
\draw [fill=lightgray] (0,0)--(6,0)--(6,6)--(0,6)--(0,0);
\node at (3,3) {$\Lambda$};
\draw [ultra thick] (6,0.1)--(6,-0.1);
\draw [ultra thick] (0.1,6)--(-0.1,6);
\draw [dotted] (6,0)--(12,12)--(0,6);
\draw [<->] (15,10)node[above]{$\lambda_{\theta_1}$}--(15,0)--(35,0)node[above]{$\lambda_{\theta_2}$};
\draw [fill=lightgray] (15,0)--(21,0)--(15,6)--(15,0);
\node at (17,1.7) {$\underline{\Lambda}$};
\draw [ultra thick] (21,0.1)--(21,-0.1);
\draw [ultra thick] (15.1,6)--(14.9,6);
\draw [dotted] (21,0)--(21,6)--(15,6);
\draw [dotted] (15,6)--(27,12)--(21,0);
\draw [<->, dashed, ultra thick, blue] (15,9)node[right]{belief after observing $H$}--(33,0)node[below]{belief after observing $L$};
\draw [ultra thick, red] (21.1,6)--(20.9,6)node[right]{$\lambda (\mu,\alpha_1^*)$};
\end{tikzpicture}
\caption{The relevant sets of likelihood ratio vectors in the example of Section 3.2, with $\overline{\Lambda}(\theta^*,\alpha_1^*,u_2)$ in the left panel, $\Lambda(\theta^*,\alpha_1^*,u_2)$ in the middle panel, and $\underline{\Lambda}(\theta^*,\alpha_1^*,u_2)$  in the right panel.}
\end{center}
\end{figure}

\subsection{Statement of Result}\label{sub3.3}
Motivated by the above example, I state a result that explores the distinctions between establishing reputations for playing pure actions and that for playing mixed actions.
For convenience, I define the likelihood ratio between every strategic type and the commitment type.
For every $\theta \in \Theta$ and $\alpha_1^* \in \mathcal{A}_1^*$,
let $\lambda_{\theta} (\mu,\alpha_1^*) \equiv \mu(\theta)/\mu(\alpha_1^*)$. Let $\lambda (\mu,\alpha_1^*) \equiv \big\{ \lambda_{\theta} (\mu,\alpha_1^*) \big\}_{\theta \in \Theta} \in \mathbb{R}_+^{m}$ be the \textit{prior likelihood ratio vector} with respect to $\alpha_1^*$.
Let $\phi_{\alpha_1^*} \in \Delta (\Theta)$ be the state distribution \textit{conditional on} player $1$ being commitment type $\alpha_1^*$.
Since $\mu$ has full support,
both $\lambda (\mu,\alpha_1^*)$ and $\phi_{\alpha_1^*}$ are well-defined.
and can be computed from player $2$'s prior belief $\mu$.

Let $\overline{\Lambda} (\theta^*,\alpha_1^*,u_2) \subset \mathbb{R}_+^m$ be the set of $m$-dimensional real vectors $\{\lambda_{\theta}\}_{\theta \in \Theta}$ that
satisfy
\begin{equation}\label{3.4}
    \big\{a_2^* (\theta^*,\alpha_1^*|u_2) \big\} = \arg \max_{a_2 \in A_2}
    \big\{  u_2(\phi_{\alpha_1^*},\alpha_1^*,a_2)+ \sum_{\theta \in \Theta } \lambda_{\theta} u_2(\theta,\alpha_1^*,a_2)
    \big\}.
\end{equation}
Intuitively, a likelihood ratio vector belongs to
$\overline{\Lambda} (\theta^*,\alpha_1^*,u_2)$ \textit{if and only if} player $2$'s best reply to $\alpha_1^*$ in state $\theta^*$ is also her best reply to $\alpha_1^*$ conditional on the event that player $1$ is either strategic or is committed to play $\alpha_1^*$.
Let
\begin{equation}\label{3.5}
    \Lambda (\theta^*,\alpha_1^*,u_2) \equiv \Big\{\lambda \in \mathbb{R}_+^m \Big| \lambda' \in \overline{\Lambda} (\theta^*,\alpha_1^*,u_2) \textrm{ for all } 0 \leq \lambda' \leq \lambda \Big\}.
\end{equation}
Theorem 1' in Pei (2020, page 2191) shows that
when all commitment actions are pure,
(\ref{3.3}) applies if $\lambda (\mu,\alpha_1^*) \in \Lambda (\theta^*,\alpha_1^*,u_2)$, and (\ref{3.3}) fails if $\lambda (\mu,\alpha_1^*) \notin \textrm{cl}\Big(\Lambda (\theta^*,\alpha_1^*,u_2) \Big)$ where $\textrm{cl}(\cdot)$ denotes the closure.
Intuitively, $\lambda (\mu,\alpha_1^*) \in \Lambda (\theta^*,\alpha_1^*,u_2)$ if and only if $a_2^*(\theta^*,\alpha_1^*|u_2)$ is player $2$'s best reply to $\alpha_1^*$ conditional on the event that player $1$ is either strategic or is committed to play $\alpha_1^*$ (i.e., $\lambda (\mu,\alpha_1^*) \in \overline{\Lambda} (\theta^*,\alpha_1^*,u_2)$), and it remains to be a best reply
when the likelihood ratio between every strategic type and commitment type $\alpha_1^*$ decreases.
When $\alpha_1^*$ is pure, every entry of the likelihood ratio vector $\lambda (\mu,\alpha_1^*)$ cannot increase when player $1$ plays $\alpha_1^*$ in every period, and the argument in Fudenberg and Levine (1989) implies that type $\theta^*$ receives at least his commitment payoff from $\alpha_1^*$ in every equilibrium.
I will extend this result when $\alpha_1^*$ is pure while other commitment actions in $\mathcal{A}_1^*$ may be mixed.

When $\alpha_1^*$ is mixed, $\lambda \in \Lambda(\theta^*,\alpha_1^*,u_2)$ is no longer sufficient since playing some actions in the support of $\alpha_1^*$ may \textit{increase} some entries of the likelihood ratio vector. I highlight such a possibility in the motivating example when the commitment action is $\alpha_1^* \equiv (1-\varepsilon) H + \varepsilon I$, under which there exist equilibria
in which the patient player's payoff is bounded below his commitment payoff from $\alpha_1^*$. This is because in period $0$,
 player $1$'s action either increases the likelihood ratio between strategic type $\theta_1$ and commitment type $\alpha_1^*$, or increases the likelihood ratio between strategic type $\theta_2$ and commitment type $\alpha_1^*$, or increases both.

I propose a sufficient condition under which
the aforementioned problem disappears.
I focus on the case in which $\Lambda(\theta^*,\alpha_1^*,u_2)$ is not empty since the case where $\Lambda(\theta^*,\alpha_1^*,u_2)=\emptyset$ is trivial. Let
\begin{equation}\label{C.5}
    \Theta_{(\alpha_1^*,\theta^*)}^b \equiv \big\{\theta \in \Theta \big|a_2^* \notin \textrm{BR}_2 (\theta,\alpha_1^*|u_2) \big\}
\end{equation}
be the set of states under which player $2$'s best reply to $\alpha_1^*$ differs from that under state $\theta^*$.
Intuitively, player $2$ has less incentive to play her best reply to $\alpha_1^*$ in state $\theta^*$ if states in
$\Theta_{(\alpha_1^*,\theta^*)}^b$ occur with high enough probability.
When $\Theta_{(\alpha_1^*,\theta^*)}^b$ is empty, the result in Fudenberg and Levine (1992) implies (\ref{3.3}), and the logic is similar to that in the private value benchmark.
If $\Theta_{(\alpha_1^*,\theta^*)}^b$ is not empty,
for every $\theta \in \Theta_{(\alpha_1^*,\theta^*)}^b$, let $\psi_{\theta}^*$ be the largest $\psi \in \mathbb{R}_+$ such that:
\begin{equation}\label{C.6}
    a_2^* \in \arg\max_{a_2 \in A_2} \Big\{ u_2(\phi_{\alpha_1^*},\alpha_1^*,a_2) +\psi u_2(\theta,\alpha_1^*,a_2) \Big\}.
\end{equation}
Intuitively,  $\psi_{\theta}^*$ is the intercept of $\Lambda(\theta^*,\alpha_1^*,u_2)$ on the axis for $\lambda_{\theta}$.
Let
\begin{equation}\label{C.7}
 \underline{\Lambda}(\theta^*,\alpha_1^*,u_2)=
 \Big\{(\lambda_{\theta})_{\theta \in \Theta} \in \mathbb{R}_+^{m} \Big|
    \sum_{\theta \in \Theta^b_{(\alpha_1^*,\theta^*)}} \lambda_{\theta}/\psi^*_{\theta}<1
    \Big\}.
\end{equation}
Since $\underline{\Lambda}(\theta^*,\alpha_1^*,u_2)$ is characterized by a linear inequality, both
$\underline{\Lambda}(\theta^*,\alpha_1^*,u_2)$ and $\mathbb{R}_+^m \backslash \underline{\Lambda}(\theta^*,\alpha_1^*,u_2)$ are convex sets.
Figure 1 depicts $\overline{\Lambda}(\theta^*,\alpha_1^*,u_2)$,
$\Lambda(\theta^*,\alpha_1^*,u_2)$, and
$\underline{\Lambda}(\theta^*,\alpha_1^*,u_2)$ in the example of Section 3.2, as well as how to obtain
 $\Lambda(\theta^*,\alpha_1^*,u_2)$
from  $\overline{\Lambda}(\theta^*,\alpha_1^*,u_2)$, and how to obtain
 $\underline{\Lambda}(\theta^*,\alpha_1^*,u_2)$ from $\Lambda(\theta^*,\alpha_1^*,u_2)$.

In order to understand why $\lambda (\mu,\alpha_1^*) \in \underline{\Lambda}(\theta^*,\alpha_1^*,u_2)$ solves the problem identified in the example, notice that
every entry of
the likelihood ratio vector is a non-negative supermartingale conditional on $\alpha_1^*$.\footnote{The likelihood ratio may not be a martingale since $\alpha_1^*$ may not have full support.}
Since $\mathbb{R}_+^m \backslash \underline{\Lambda}(\theta^*,\alpha_1^*,u_2)$ is convex, if the prior likelihood ratio vector belongs to
$\underline{\Lambda}(\theta^*,\alpha_1^*,u_2)$, then
there exists at least one pure action $a_1$ in the support of $\alpha_1^*$ such that the posterior likelihood ratio vector belongs to $\underline{\Lambda}(\theta^*,\alpha_1^*,u_2)$ after observing $a_1$.
This is because otherwise, the posterior likelihood ratio after observing every action in the support of $\alpha_1^*$ does not belong to
$\underline{\Lambda}(\theta^*,\alpha_1^*,u_2)$, and since $\mathbb{R}_+^m \backslash \underline{\Lambda}(\theta^*,\alpha_1^*,u_2)$ is convex,
the prior likelihood ratio also does not belong to
$\underline{\Lambda}(\theta^*,\alpha_1^*,u_2)$, which leads to a contradiction.

By contrast, $\mathbb{R}_+^m \backslash \Lambda(\theta^*,\alpha_1^*,u_2)$ is not necessarily convex, which implies the possibility that the prior likelihood ratio belongs to $\Lambda(\theta^*,\alpha_1^*,u_2)$
under which player $2$ has an incentive to play the desired best reply to $\alpha_1^*$, but no matter which action player $1$ takes in the support of $\alpha_1^*$, player $2$'s posterior likelihood ratio does not belong to $\Lambda(\theta^*,\alpha_1^*,u_2)$, and hence, player 2 has a strict incentive not to play $a_2^* (\theta^* , \alpha_1^*|u_2)$.
\begin{Theorem}\label{Theorem1}
For every pure commitment action $\alpha_1^* \in \mathcal{A}_1^*$
and every $\theta^* \in \Theta$,
\begin{enumerate}
\item[1.] If $\lambda (\mu,\alpha_1^*) \in \Lambda (\theta^*,\alpha_1^*,u_2)$, then $\liminf_{\delta \rightarrow 1} \underline{v}_{\theta^*}(\delta,\mu, u_1,u_2) \geq v_{\theta^*}(\alpha_1^*, u_1,u_2)$ for every $u_1$, i.e.,
    strategic type $\theta^*$ can guarantee his complete information commitment payoff from $\alpha_1^*$ regardless of his stage-game payoff function when he is sufficiently patient.
\item[2.] If $\lambda  (\mu,\alpha_1^*)$ does not belong to the closure of $\Lambda (\theta^*,\alpha_1^*,u_2)$ and $\textrm{BR}_2(\phi_{\alpha_1^*},\alpha_1^*|u_2)$ is a singleton, then there exists $u_1$ such that $\limsup_{\delta \rightarrow 1} \underline{v}_{\theta^*}(\delta,\mu, u_1,u_2) <v_{\theta^*}(\alpha_1^*, u_1,u_2)$, i.e., there exist a stage-game payoff function for player $1$ and an equilibrium in which strategic type $\theta^*$'s payoff is bounded below his complete information commitment payoff from $\alpha_1^*$ no matter how patient he is.
\end{enumerate}
For every nontrivially mixed commitment action $\alpha_1^* \in \mathcal{A}_1^*$
and every $\theta^* \in \Theta$,
\begin{enumerate}
\item[3.] If $\lambda  (\mu,\alpha_1^*) \in \underline{\Lambda}(\theta^*,\alpha_1^*,u_2)$, then $\liminf_{\delta \rightarrow 1} \underline{v}_{\theta^*}(\delta,\mu, u_1,u_2) \geq v_{\theta^*}(\alpha_1^*, u_1,u_2)$ for every $u_1$.
\item[4.] If $\lambda  (\mu,\alpha_1^*)$ does not belong to the closure of $\underline{\Lambda} (\theta^*,\alpha_1^*,u_2)$, $\textrm{BR}_2(\phi_{\alpha_1^*},\alpha_1^* |u_2)$ is a singleton and $\alpha_1^*$ does not belong to the convex hull of $\mathcal{A}_1^* \Big\backslash \{\alpha_1^*\}$, then there exists $u_1$ such that $\limsup_{\delta \rightarrow 1} \underline{v}_{\theta^*}(\delta,\mu, u_1,u_2) < v_{\theta^*}(\alpha_1^*, u_1,u_2)$.
\end{enumerate}
\end{Theorem}
My proof consists of two parts. I establish a lower bound on the patient player's equilibrium payoff in Section \ref{sec4} and Appendix A, which shows statements 1 and 3.
I construct equilibria in which the patient player's payoff is bounded below his commitment payoff
in Section \ref{sec5} and the Online Appendix, which shows statements 2 and 4. The main technical contribution is the proof of a learning result stated as Proposition \ref{Prop1}, which is a key step to establish statement 3 and is also potable to study games with imperfect monitoring (Section \ref{sec6}).

Theorem \ref{Theorem1} has two implications. First, it
points out the failure of reputation effects
in repeated incomplete information games with nontrivial interdependent values.
According to this interpretation,
the interdependent value reputation model is obtained by perturbing
a repeated incomplete information game with a small probability of commitment types. When
every commitment type is \textit{arbitrarily unlikely} relative to every strategic type and player $2$'s best reply to $\alpha_1^*$ depends on the state,
the prior likelihood ratio vector $\lambda (\mu,\alpha_1^*)$ does not belong to the closures of $\Lambda (\theta,\alpha_1^*,u_2)$ and $\underline{\Lambda}(\theta,\alpha_1^*,u_2)$ for any $\theta \in \Theta$.



Theorem \ref{Theorem1} also evaluates the robustness of reputation effects in private value reputation games against \textit{interdependent value perturbations}. Under this interpretation, a private value reputation game is perturbed with a small probability of \textit{other strategic types}. Such a perturbation captures situations such as buyers facing uncertainty about the safety or durability of the seller's products, which the seller knows more about. My sufficient conditions are satisfied when the short-run players' doubt on their own payoffs is sufficiently small.

\section{Establishing Lower Bounds on Player $1$'s Equilibrium Payoff}\label{sec4}
My proof of the first statement resembles that of Theorem 1' in Pei (2020). If $\alpha_1^*$ is pure and
$\lambda (\mu,\alpha_1^*) \in \Lambda (\theta^*,\alpha_1^*,u_2)$, then the patient player can secure his commitment payoff by playing $\alpha_1^*$ in every period. This is because the posterior probability of other commitment types vanishes to zero exponentially under such a strategy and will have negligible impact on player $2$'s best reply. I omit the details in order to avoid repetition.

The substantial difference arises when $\alpha_1^*$ is nontrivially mixed. I construct a strategy for strategic type $\theta^*$ under which he can guarantee his complete information commitment payoff from $\alpha_1^*$. For every $\psi \equiv (\psi_{\theta})_{\theta \in \Theta} \in \mathbb{R}_+^{m}$ and $\chi > 0$, let
\begin{equation}\label{C.8}
    \underline{\Lambda}(\psi,\chi) \equiv
    \Big\{ ( \widetilde{\lambda}_{\theta})_{\theta \in \Theta} \in \mathbb{R}_+^{m} \Big|
    \sum_{\theta \in \Theta} \widetilde{\lambda}_{\theta}/\psi_{\theta} < \chi
    \Big\}.
\end{equation}
Let $\mu(h^t)$ be player $2$'s posterior belief at $h^t$. I write $\lambda(h^t)$ in short for $\lambda (\mu(h^t),\alpha_1^*)$, which is the likelihood ratio vector with respect to commitment action $\alpha_1^*$ at $h^t$.
Let $h^{\infty}$ be an infinite history.
Let $\overline{A_1} \equiv \textrm{supp}(\alpha_1^*)$.
For every $\sigma_{\theta}: \mathcal{H} \rightarrow \Delta(A_1)$ and $\sigma_2: \mathcal{H} \rightarrow \Delta(A_2)$, let
$\mathcal{P}^{(\sigma_{\theta},\sigma_2)}$ be the probability measure over $\mathcal{H}$ induced by $(\sigma_{\theta},\sigma_2)$, let $\mathcal{H}^{(\sigma_{\theta},\sigma_2)}$ be the set of histories that occur with positive probability under $\mathcal{P}^{(\sigma_{\theta},\sigma_2)}$, and let $\mathbb{E}^{(\sigma_{\theta},\sigma_2)}$ be the expectation induced by $\mathcal{P}^{(\sigma_{\theta},\sigma_2)}$.
\begin{Proposition}\label{Prop1}
Suppose $\lambda \in  \underline{\Lambda}(\psi,\chi)$. For every $\epsilon>0$, there exist $T \in \mathbb{N}$ and $\underline{\delta} \in (0,1)$ such that for every $\delta > \underline{\delta}$ and every equilibrium $\sigma \equiv ((\sigma_{\theta})_{\theta \in \Theta}, \sigma_2)$, we can find a deviation $\widetilde{\sigma}_{\theta}: \mathcal{H} \rightarrow \Delta (\overline{A}_1)$ and a continuous function $\beta (\delta)$ satisfying $\lim_{\delta \rightarrow 1} \beta(\delta)=0$ such that:
  \begin{equation}\label{C.9}
  \lambda (h^t) \in \underline{\Lambda}(\psi,\chi+\epsilon)
 \quad \textrm{for every} \quad h^t \in \mathcal{H}^{(\widetilde{\sigma}_{\theta},\sigma_2)},
  \end{equation}
  \begin{equation}\label{C.10}
 \mathcal{P}^{(\widetilde{\sigma}_{\theta},\sigma_2)}  \Big(  \Big|
    \sum_{t=0}^{\infty} (1-\delta)\delta^t \mathbf{1}\{h^{\infty}_t=a_1\}-\alpha_1^*(a_1)
    \Big|<\epsilon \textrm{ for every } a_1 \in A_1 \Big) >1- \beta (\delta),
  \end{equation}
  \begin{equation}\label{C.11}
   \mathbb{E}^{(\widetilde{\sigma}_{\theta},\sigma_2)} \Big[
    \# \Big\{t  \in \mathbb{N}\Big| ||\alpha_1^* -\alpha_1(\cdot|h^t) ||>\epsilon \Big\}
    \Big]
   <T.
  \end{equation}
\end{Proposition}
Recall the definition of $\underline{\Lambda}(\theta^*,\alpha_1^*,u_2)$ in (\ref{C.7}). One can verify that $\underline{\Lambda} (\theta^*,\alpha_1^*,u_2)$ coincides with $\underline{\Lambda}(\psi,\chi)$ when $\chi=1$,
$\psi_{\theta} \equiv \psi^*_{\theta}$ for every $\theta \in \Theta_{(\alpha_1^*,\theta^*)}^b$, and $\psi_{\theta} \equiv +\infty$ for every
$\theta \notin \Theta_{(\alpha_1^*,\theta^*)}^b$. Let $\epsilon \equiv \frac{1}{2} \Big(1- \sum_{\theta \in \Theta} \frac{\lambda_{\theta}(\mu,\alpha_1^*)}{\psi_{\theta}}\Big)$, which is strictly positive when $\lambda (\mu,\alpha_1^*) \in \underline{\Lambda}(\theta^*,\alpha_1^*,u_2)$.
Proposition \ref{Prop1} implies Corollary 1:
\begin{Corollary}
If $\lambda (\mu,\alpha_1^*) \in \underline{\Lambda}(\theta^*,\alpha_1^*,u_2)$ and $\delta$ is large,
then for every equilibrium $\sigma$,
there exists a deviation for strategic type $\theta^*$, denoted by $\widetilde{\sigma}_{\theta^*} : \mathcal{H} \rightarrow \Delta (\overline{A}_1)$ such that when player $1$ uses $\widetilde{\sigma}_{\theta^*}$ and player $2$s use their equilibrium strategy,
\begin{itemize}
  \item[1.] With probability $1$, player 2's posterior likelihood ratio vector in every period belongs to
  $\Lambda (\psi, 1-\epsilon)$.
  \item[2.] With probability close to $1$, the discounted frequency of every $a_1 \in A_1$ is
 approximately $\alpha_1^*(a_1)$.
  \item[3.] In all but a bounded number of periods, player 2's prediction about player 1's action is close to $\alpha_1^*$
\end{itemize}
\end{Corollary}
I show Proposition \ref{Prop1} in the rest of this section in three steps, which in turn implies Corollary 1.
The remaining steps of the proof after establishing Corollary 1
is relegated to Appendix \ref{secA}.

\paragraph{Step 1:} Let $\mathcal{P}^{(\alpha_1^*,\sigma_2)}$ be the probability measure over $\mathcal{H}$
when player $1$ plays $\alpha_1^*$ in every period and player $2$ plays according to $\sigma_2$.
Let $\chi(h^t) \equiv \sum_{i=1}^m \lambda_i(h^t)/\psi_i$.
By definition,
$\lambda \in \underline{\Lambda}(\psi,\chi)$ if and only if $\chi(h^0) < \chi$.
Let $\{\mathcal{F}^t\}_{t\in \mathbb{N}}$ be the filtration induced by the public history.
Since $\{\lambda_i(h^t),\mathcal{P}^{(\alpha_1^*,\sigma_2)},\mathcal{F}^t \}_{t \in \mathbb{N}}$ is a non-negative supermartingale for every $i \in \{1,2,...,m\}$,
$\{\chi_t,\mathcal{P}^{(\alpha_1^*,\sigma_2)},\mathcal{F}^t \}_{t \in \mathbb{N}}$
is also a non-negative supermartingale. For every $a<b$, let $U(a,b)$ be the number of upcrossings from $a$ to $b$.
The Doob's Upcrossing Inequality implies:
\begin{equation}\label{C.13}
    \mathcal{P}^{(\alpha_1^*,\sigma_2)} \Big\{
    U(\chi,\chi+\epsilon) =0
    \Big\} \geq \frac{\epsilon}{\chi+\epsilon}.
\end{equation}
Let $\widetilde{\mathcal{H}}^{\infty}$ be the set of histories such that $\chi_t < \chi+\epsilon$ for every $t \in \mathbb{N}$. According to (\ref{C.13}), $\widetilde{\mathcal{H}}^{\infty}$ occurs with probability at least $\frac{\epsilon}{\chi+\epsilon}$ under probability measure $\mathcal{P}^{(\alpha_1^*,\sigma_2)}$.

Let $\widetilde{\mathcal{P}}$ be a probability measure defined as $\widetilde{\mathcal{P}} (E) \equiv \mathcal{P}^{(\alpha_1^*,\sigma_2)}(E \cap \widetilde{\mathcal{H}}^{\infty})/ \mathcal{P}^{(\alpha_1^*,\sigma_2)} (\widetilde{\mathcal{H}}^{\infty})$.
I construct a strategy $\widetilde{\sigma}_{\theta}$ such that when player $1$ uses
$\widetilde{\sigma}_{\theta}$ and player $2$s use their equilibrium strategy, the induced probability measure over histories is
$\widetilde{\mathcal{P}}$.
For every $h^t$ such that $\chi(h^t)< \chi+\epsilon$, let $A_1 (h^t) \subset \textrm{supp}(\alpha_1^*)$ be such that $a_1 \in A_1 (h^t)$ if and only if $\chi(h^t,a_1)< \chi+\epsilon$. The set $A_1(h^t)$ is not empty since $\chi(h^t)< \chi+\epsilon$ and
$\{\chi_t,\mathcal{P}^{(\alpha_1^*,\sigma_2)},\mathcal{F}^t \}_{t \in \mathbb{N}}$ is a supermartingale, and moreover, $h^t \in \widetilde{\mathcal{H}}^{\infty}$ if and only if for every $s <t$, $h^s \in \widetilde{\mathcal{H}}^{\infty}$ and player $1$'s action in period $s$ belongs to $A_1(h^s)$.
Let $\widetilde{\mathcal{P}}(\cdot|h^t)$ be the probability measure induced by  $\widetilde{\mathcal{P}}$ conditional on the history being $h^t$, which is well-defined for every $h^t \in \widetilde{\mathcal{H}}^{\infty}$. Suppose
$\widetilde{\sigma}_{\theta}$ is such that at every $h^t$ satisfying $\chi(h^t)< \chi+\epsilon$, player $1$ plays $a_1$ with zero probability if $a_1 \notin A_1(h^t)$, and plays $a_1$ with probability $\widetilde{\mathcal{P}}(a_{1,t}=a_1|h^t)$
if $a_1 \in A_1 (h^t)$; at every $h^t$ such that $\chi(h^t) \geq \chi+\epsilon$,
$\widetilde{\sigma}_{\theta}$ can be arbitrary.
By construction, $\widetilde{\sigma}_{\theta}$ induces probability measure $\widetilde{\mathcal{P}}$.

\paragraph{Step 2:} I show that when $\delta$ is close enough to $1$, there exists a subset of $\mathcal{H}^{\infty}$ that occurs with probability close to $1$ under probability measure $\mathcal{P}^{(\alpha_1^*,\sigma_2)}$, such that
the discounted frequency of every $a_1 \in A_1$ is close to $\alpha_1^* (a_1)$.
For every $a_1 \in \overline{A}_1$,
let $\{X_t\}$ be a sequence of i.i.d. random variables such that:
\begin{displaymath}
X_t =\left\{ \begin{array}{ll}
1 & \textrm{ when } a_{1,t}=a_1 \\
0 & \textrm{ otherwise }.
\end{array} \right.
\end{displaymath}
Under probability measure $\mathcal{P}^{(\alpha_1^*,\sigma_2)}$, $X_t=1$ occurs with probability $\alpha_1^*(a_1)$. Let $n \equiv |A_1|$. Lemma 4.1 is implied by the Central Limit Theorem for triangular sequences (Chung 1974), with proof in Appendix \ref{secB}.
\begin{Lemma}\label{L4.2}
For every $\eta>0$, there exists $\overline{\delta} \in (0,1)$, such that for all $\delta \in (\overline{\delta},1)$,
\begin{equation}\label{C.14}
   \lim \sup_{\delta \rightarrow 1} \mathcal{P}^{(\alpha_1^*,\sigma_2)} \Big(
     \Big|  \sum_{t=0}^{+\infty} (1-\delta) \delta^t X_t -\alpha_1^*(a_1) \Big| \geq \eta
    \Big)\leq \frac{\eta}{n}.
\end{equation}
\end{Lemma}
 According to Lemma \ref{L4.2}, for every $a_1 \in A_1$ and $\eta>0$, there exists $\overline{\delta} \in (0,1)$, such that for every $\delta>\overline{\delta}$, there exists $\mathcal{H}_{\eta,a_1}^{\infty}(\delta) \subset \mathcal{H}^{\infty}$, such that
the discounted frequency of $a_1$ is $\eta$-close to $\alpha_1^*(a_1)$ for every $h^{\infty} \in \mathcal{H}^{\infty}_{\eta,a_1}(\delta)$, and $\mathcal{P}^{(\alpha_1^*,\sigma_2)} (\mathcal{H}_{\eta,a_1}^{\infty}(\delta)) \geq 1-\eta/n$. Let $\mathcal{H}_{\eta}^{\infty}(\delta) \equiv \bigcap_{a_1 \in A_1} \mathcal{H}^{\infty}_{\eta,a_1}(\delta)$, we have
\begin{equation}\label{C.16}
   \mathcal{P}^{(\alpha_1^*,\sigma_2)}(\mathcal{H}^{\infty}_{\eta}(\delta)) \geq 1-\eta.
\end{equation}

\paragraph{Step 3:}
Recall that $\widetilde{\mathcal{H}}^{\infty}$ is the set of histories such that $\chi_t < \chi+\epsilon$ for every $t \in \mathbb{N}$, which occurs with probability at least $\frac{\epsilon}{\chi+\epsilon}$. Therefore, the probability of  $\widehat{\mathcal{H}}^{\infty} \equiv \widetilde{\mathcal{H}}^{\infty} \bigcap \mathcal{H}_{\eta}^{\infty}(\delta)$
conditional on $\widetilde{\mathcal{H}}^{\infty}$ is at least
$1-\frac{\eta (\chi +\epsilon)}{\epsilon}$.
Intuitively, $\widehat{\mathcal{H}}^{\infty}$ is the event in which $\chi_t < \chi+\epsilon$ for every $t \in \mathbb{N}$ and the discounted frequency of every player $1$'s pure action is $\eta$-close to its probability in $\alpha_1^*$.
Since $\eta$ is arbitrarily close to $0$ as $\delta \rightarrow 1$, 
$1-\frac{\eta (\chi +\epsilon)}{\epsilon}$ can be arbitrarily close to $1$, which means that
the probability that the discounted frequency of every action being close to its probability in the mixed commitment action is arbitrarily close to $1$ conditional on $\widetilde{\mathcal{H}}^{\infty}$.

Let $d(\cdot \| \cdot)$ denote the  Kullback-Leibler divergence
 between two distributions.
Gossner (2011)'s result implies that:
\begin{equation}\label{C.19}
  \mathbb{E}^{(\alpha_1^*,\sigma_2)} \Big[   \sum_{\tau=0}^{+\infty} d(\alpha_1^*||\alpha_1(\cdot|h^{\tau}))
    \Big] \leq -\log \mu(\alpha_1^*).
\end{equation}
Since the Kullback-Leibler divergence must be non-negative,
Markov Inequality implies that:
\begin{equation}\label{C.20}
  \mathbb{E}^{(\alpha_1^*,\sigma_2)} \Big[   \sum_{\tau=0}^{+\infty} d(\alpha^*||\alpha(\cdot|h^{\tau})) \Big| \widetilde{\mathcal{H}}^{\infty}
    \Big] \leq -\frac{(\chi+\epsilon)\log \mu(\alpha_1^*)}{\epsilon}.
\end{equation}
Recall that $\widetilde{\sigma}_{\theta}$ is strategic-type player $1$'s strategy that induces probability measure
$\widetilde{\mathcal{P}}$, i.e., the probability measure such that $\widetilde{\mathcal{P}} (E) \equiv \mathcal{P}^{(\alpha_1^*,\sigma_2)}(E \cap \widetilde{\mathcal{H}}^{\infty})/ \mathcal{P}^{(\alpha_1^*,\sigma_2)} (\widetilde{\mathcal{H}}^{\infty})$.
If a strategic-type player $1$ deviates to $\widetilde{\sigma}_{\theta}$, then
the expected number of periods in which $d(\alpha_1^*||\alpha(\cdot|h^t))>\epsilon^2/2$ is at most:
\begin{equation}\label{C.21}
    T \equiv  \Big\lceil -\frac{2(\chi+\epsilon)\log \mu(\alpha_1^*)}{\epsilon^3} \Big\rceil.
\end{equation}
The Pinsker's inequality implies that the expected number of periods in which $||\alpha_1^*-\alpha(\cdot|h^t)||>\epsilon$ is
at most $T$. The three steps together imply Proposition \ref{Prop1}.

\paragraph{Summary of Remaining Steps:} Proposition \ref{Prop1} and Corollary 1
do not directly imply that type $\theta^*$ can guarantee payoff $v_{\theta^*}(\alpha_1^*,u_1,u_2)$ for every $u_1$ in every equilibrium.
This is because due to the potential correlation between player $1$'s action and the state $\theta$,
player $2$s may not have incentives to play $a_2^*$
despite $\lambda (\mu,\alpha_1^*)$ belongs to $\underline{\Lambda} (\theta^*,\alpha_1^*,u_2)$ and is bounded away from the boundary of
$\underline{\Lambda} (\theta^*,\alpha_1^*,u_2)$,
and player $1$'s average action is close to $\alpha_1^*$. For a summary of the proof after establishing Corollary 1:
\begin{enumerate}
  \item Suppose all entries of $\lambda(\mu,\alpha_1^*)$ except for at most one is sufficiently small,
  then player $2$ has a strict incentive to play $a_2^*$ when player $1$'s average action is close to $\alpha_1^*$. Let $\Lambda^0$ be the set of type distributions with this feature. One can then directly apply Corollary 1 to establish inequality (\ref{3.3}).
  \item If player $1$'s average action is close to $\alpha_1^*$ but player $2$ does not have a strict incentive to play $a_2^*$, then
  different types of player $1$'s actions at that history must be significantly different. This implies that player $1$'s action at that history must be informative about his type, in which case he can pick a particular action that induces player $2$ to learn.
      I show that for every $\lambda (\mu,\alpha_1^*) \in \underline{\Lambda} (\theta^*,\alpha_1^*,u_2)$, there exists an integer $K(\lambda)$ and a strategy for type $\theta^*$ such that if type $\theta^*$ follows this strategy, then after at most $K(\lambda)$ such periods, player $2$'s belief about his type belongs to $\Lambda^0$, which concludes the proof.
\end{enumerate}

\section{Constructing Low Payoff Equilibria}\label{sec5}
This section provides a constructive proof for the second statement of Theorem \ref{Theorem1}. I replace $\alpha_1^*$ with $a_1^*$ since it is pure.
When all actions in $\mathcal{A}_1^*$ are pure and player $1$'s stage-game payoff function is $u_1(\theta,a_1,a_2) \equiv \mathbf{1}\{\theta=\theta^*,a_1=a_1^*,a_2=a_2^*\}$,
Pei (2020) constructs an equilibrium in which
all strategic types in $\Theta_{(a_1^*,\theta^*)}^b$ play $a_1^*$ in every period, the other strategic types play $a_1^*$ in every period, and
the patient player's payoff is $0$.

Such a construction \textit{does not} work when there are commitment types other than $a_1^*$ who  play mixed strategies. Consider an example in which $\Theta=\{\theta,\widetilde{\theta}\}$, $\tilde{\theta} \in \Theta^b_{(a_1^*,\theta)}$,
$\mathcal{A}_1^*=\{a_1^*,\alpha_1'\}$,
$\alpha_1'$ is non-trivially mixed, attaching positive probability to $a_1^*$, with
$\{a_2^*\}=\textrm{BR}_2(\phi_{a_1^*},a_1^*|u_2)=\textrm{BR}_2(\phi_{\alpha_1'},\alpha_1'|u_2)$.
If type $\widetilde{\theta}$ plays $a_1^*$ in every period,
then type $\theta$ can obtain a payoff arbitrarily close to $1$ by
playing $a_1 \in \textrm{supp}(\alpha_1') \backslash \{a_1^*\}$ in period $0$ and $a_1^*$ in every subsequent period. The reason is that after observing $a_1$ in period $0$, player $2$s attach probability $1$ to commitment type $\alpha_1'$ and have a strict incentive to play $a_2^*$.

In order
to overcome this challenge, I construct an equilibrium in which the strategic types in $\Theta^b_{(\theta^*,a_1^*)}$ play \textit{non-stationary strategies}.
In the above example, type $\widetilde{\theta}$ plays $a_1^*$ in every period with probability $p \in (0,1)$ and plays non-stationary strategy $\sigma(\alpha_1')$ with probability $1-p$, with $p$ being large enough such that $\lambda_1$ is bounded away from $\overline{\Lambda}$ after $a_1^*$ is played in period $0$.
Strategy
$\sigma (\alpha_1')$ is described as follows:
\begin{itemize}
\item Play $\alpha_1'$ at histories that are consistent with type $\theta$'s equilibrium strategy.
\item Otherwise, play a completely mixed action $\widehat{\alpha}'_1$ that attaches higher probability to $a_1^*$ compared to $\alpha_1'$.
\end{itemize}
\paragraph{Step 1:} I show that when $\lambda (\mu,a_1^*) \notin \textrm{clo}\Big(\Lambda(\theta^*,a_1^*,u_2)\Big)$,
there exist
$a_2 \neq a_2^*$ and $\lambda' \equiv \{\lambda_{\theta}'\}_{\theta \in \Theta}$, such that first,
$0 \leq \lambda' \leq \lambda(\mu,a_1^*)$ and $\lambda_{\theta^*}'=0$, second,
\begin{equation}\label{B.1}
    \sum_{\theta \in \Theta} \lambda_{\theta}' \Big(u_2(\theta,a_1^*,a_2)
-u_2(\theta,a_1^*,a_2^*) \Big) >0,
\end{equation}
and third,
\begin{equation}\label{B.2}
u_2(\phi_{a_1^*},a_1^*,a_2) - u_2(\phi_{a_1^*},a_1^*,a_2^*) +    \sum_{\theta \in \Theta} \lambda_{\theta}' \Big(u_2(\theta,a_1^*,a_2)
-u_2(\theta,a_1^*,a_2^*) \Big) >0.
\end{equation}
According to the definition of $\Lambda (\theta^*,a_1^*,u_2)$ in (\ref{3.5}), there exists $\lambda'' \equiv \{\lambda_{\theta}''\}_{\theta \in \Theta}$ such that
$0 \leq \lambda'' \leq \lambda(\mu,a_1^*)$, and
\begin{equation*}
    a_2^* \notin \arg\max_{a_2 \in A_2} \Big\{ u_2(\phi_{a_1^*},a_1^*,a_2')+ \sum_{\theta \in \Theta } \lambda_{\theta}'' u_2(\theta,a_1^*,a_2') \Big\}.
\end{equation*}
Let $\lambda'\in \mathbb{R}_+^m$ be such that $\lambda_{\theta^*}'\equiv 0$, and $\lambda_{\theta}'\equiv \lambda_{\theta}''$ for all $\theta \neq \theta^*$. Since $\{a_2^*\} =\textrm{BR}_2(\theta^*,a_1^*|u_2)$,
there exists
$a_2' \neq a_2^*$:
\begin{equation*}
     u_2(\phi_{a_1^*},a_1^*,a_2')+ \sum_{\theta \in \Theta } \lambda_{\theta}' u_2(\theta,a_1^*,a_2')
     >
     u_2(\phi_{a_1^*},a_1^*,a_2^*)+ \sum_{\theta \in \Theta } \lambda_{\theta}' u_2(\theta,a_1^*,a_2^*).
\end{equation*}
If the unique element in $\textrm{BR}_2(\phi_{a_1^*},a_1^*|u_2)$ is $a_2^*$, then (\ref{B.1}) and (\ref{B.2}) hold for $a_2=a_2'$. If the unique element in
$\textrm{BR}_2(\phi_{a_1^*},a_1^*|u_2)$ is $a_2'' \neq a_2^*$,
then there exists $\theta' \in \Theta$ such that $u_2(\theta',a_1^*,a_2'') > u_2(\theta',a_1^*,a_2^*)$.
Let $\lambda'  \equiv (\lambda_{\theta}')_{\theta \in \Theta}\in \mathbb{R}_+^m$ be defined as:
$\lambda_{\theta'}' \equiv \lambda_{\theta'}$, and $\lambda_{\theta}' \equiv 0$ for all $\theta \neq \theta'$, then (\ref{B.1}) and (\ref{B.2}) hold for $\lambda'$ and $a_2=a_2''$.


\paragraph{Step 2:} Let
\begin{equation}\label{B.3}
    u_1(\theta,a_1,a_2) \equiv \mathbf{1}\{\theta=\theta^*,a_1=a_1^*,a_2=a_2^*\}.
\end{equation}
By definition, $v_{\theta^*}(a_1^*, u_1,u_2)=1$. I describe players' equilibrium strategies.
On the equilibrium path, strategic type $\theta^*$ plays a different pure action in each period from period $0$ to $|A_1|-1$.
Starting from period $|A_1|$,
he plays $a_1^*$ for $k^* \in \mathbb{N}$ periods and then some prespecified $a_1 \neq a_1^*$ in the $k^*+1$th period.
His on-path behavior rotates every $k^*+1$ periods.
I will specify the value of integer $k^*$ by the end of step 3.

I construct $\lambda' \in \mathbb{R}_+^m$ and $a_2' \neq a_2^*$ according to Step 1.
Inequality (\ref{B.2}) implies the existence of $\epsilon>0$ such that:
\begin{equation}\label{B.3}
u_2(\phi_{a_1^*},a_1^*,a_2') - u_2(\phi_{a_1^*},a_1^*,a_2^*) +   (1-\epsilon)  \sum_{\theta \in \Theta} \lambda_{\theta}' \Big(u_2(\theta,a_1^*,a_2')
-u_2(\theta,a_1^*,a_2^*) \Big) >0.
\end{equation}
For every $\widetilde{\theta} \neq \theta^*$,
with probability $\Big(\lambda_{\widetilde{\theta}} - \lambda_{\widetilde{\theta}}' \Big) \Big/ \lambda_{\widetilde{\theta}}$,
strategic type $\widetilde{\theta}$ plays $a_1' \neq a_1^*$ in every period; with probability $(1-\epsilon) \lambda_{\widetilde{\theta}}' / \lambda_{\widetilde{\theta}}$, strategic type $\widetilde{\theta}$ plays $a_1^*$ in every period.
For every $\alpha_1 \in \mathcal{A}_1^*$ that is nontrivially mixed, strategic type $\widetilde{\theta}$ plays strategy $\sigma_{\alpha_1}$ with probability
$\frac{\epsilon}{k} \lambda_{\widetilde{\theta}}'\Big/ \lambda_{\widetilde{\theta}}$,
with $k \in \mathbb{N}$ being the number of nontrivially mixed commitment actions in $\mathcal{A}_1^*$
and $\sigma_{\alpha_1}$ will be specified in the next paragraph.
If $k=0$, then one can set $\epsilon=0$.

Next, I describe strategy $\sigma_{\alpha_1}$.
If $h^t$ occurs with positive probability under
strategic type $\theta^*$'s equilibrium strategy, then $\sigma_{\alpha_1} (h^t) = \alpha_1$.
If $h^t$ occurs with zero probability under
strategic type $\theta^*$'s equilibrium strategy,
then $\sigma_{\alpha_1} (h^t) = \widehat{\alpha}_1$ where
\begin{equation}
\widehat{\alpha}_1(\alpha_1)\equiv (1-\frac{\eta}{2}) a_1^*+\frac{\eta}{2} \widetilde{\alpha}_1(\alpha_1)
\end{equation}
and
\begin{equation}\label{B.6}
\widetilde{\alpha}_1(\alpha_1)[a_1] \equiv \left\{ \begin{array}{ll}
0 & \textrm{ when } a_{1}=a_1^* \\
\alpha_1(a_1)/(1-\alpha_1(a_1^*)) & \textrm{ otherwise }.
\end{array} \right.
\end{equation}
Since $\mathcal{A}_1^*$ is a finite set,
there exists $\eta>0$ such that
$\max_{\alpha_1 \in \mathcal{A}_1^* \backslash\{a_1^*\}} \alpha_1(a_1^*)<1-\eta$.
According to (\ref{B.1}), for every $\alpha_1' \in \Delta(A_1)$ with $\alpha_1'(a_1^*)\geq 1-\eta$, we have:
\begin{equation}\label{B.7}
  \sum_{\theta \in \Theta} \lambda_{\theta}' u_2(\theta,\alpha_1',a_2')   >\sum_{\theta \in \Theta}  \lambda_{\theta}' u_2(\theta,\alpha_1',a_2^*).
\end{equation}

\paragraph{Step 3:} I verify type $\theta^*$'s incentive constraints by deriving a \textit{uniform upper bound} on his continuation payoff \textit{after his first deviation}.
For every $\alpha_1 \in\mathcal{A}_1^*$, let $\mu_{t}(\theta(\alpha_1))$ be the probability that player $1$ is strategic and follows strategy $\sigma_{\alpha_1}$. Let $\beta_t(\alpha_1) \equiv \mu_{t}(\theta(\alpha_1))/\mu_{t}(\alpha_1)$.
The value of $\beta_t(\alpha_1)$ equals $\beta_0(\alpha_1)$
at period $t$ histories that occur with positive probability under
type $\theta^*$'s equilibrium strategy.

Next, consider histories that occur with zero probability under
strategic type $\theta^*$'s equilibrium strategy.
Since $\max_{\alpha_1 \in \mathcal{A}_1^* \backslash \{a_1^*\}} \alpha_1(a_1^*)<1-\eta$, we know that when $a_1^*$ is observed in period $t$, $\beta_{t+1}(\alpha_1) \geq \frac{1-\eta/2}{1-\eta}\beta_t(\alpha_1)$
for every $\alpha_1 \in \mathcal{A}_1^* \backslash \{a_1^*\}$.
Let $\kappa \equiv 1-\min_{\alpha_1 \in \mathcal{A}_1^* \backslash \{a_1^*\}} \alpha_1(a_1^*)$. If $a_1 \neq a_1^*$ is observed in period $t$, then the definition of $\widetilde{\alpha}_1(\alpha_1)$ implies that
\begin{equation}
\beta_{t+1}(\alpha_1) \geq \frac{\eta}{2\kappa}\beta_t(\alpha_1).
\end{equation}
Let $\overline{k} \equiv \Big\lceil
    \log \frac{2\kappa}{\eta} \Big/ \log \frac{1-\eta/2}{1-\eta}
    \Big\rceil$. For every $\alpha_1 \in \mathcal{A}_1^*$, let $\overline{\beta}(\alpha_1)$ be the smallest $\beta \in \mathbb{R}_+$ such that:
\begin{equation}
u_2(\phi_{\alpha_1},\alpha_1,a_2')+\beta
\sum_{\theta \in \Theta} \lambda_{\theta}' u_2(\theta,\widehat{\alpha}_1(\alpha_1),a_2')
\geq u_2(\phi_{\alpha_1},\alpha_1,a_2^*)+\beta
\sum_{\theta \in \Theta} \lambda_{\theta}' u_2(\theta,\widehat{\alpha_1}(\alpha_1),a_2^*)
\end{equation}
Let $\overline{\beta} \equiv 2\max_{\alpha_1 \in \mathcal{A}_1^* \backslash \{a^*\}} \overline{\beta}(\alpha_1)$ and $\underline{\beta} \equiv \min_{\alpha_1 \in \mathcal{A}_1^* \backslash \{a^*\}} \frac{\mu(\theta(\alpha_1))}{\mu(\alpha_1)}$. Let
    $T_1 \equiv \Big\lceil
    \log \frac{\overline{\beta}}{\underline{\beta}} \Big/ \log \frac{1-\eta/2}{1-\eta}
    \Big\rceil$.
At any history right after type $\theta^*$'s first deviation,
$\beta_t(\alpha_1) \geq \underline{\beta}$ for all $\alpha_1 \in \mathcal{A}_1^*$. After player $2$ observes $a_1^*$ for $T_1$ consecutive periods, $a_2^*$ is strictly dominated by $a_2'$ until some $a_1' \neq a_1^*$ is observed. Moreover, every time player $1$ plays some $a_1' \neq a_1^*$, he can induce outcome $(a_1^*,a_2^*)$ for at most $\overline{k}$ consecutive periods before $a_2^*$ is strictly dominated by $a_2'$ again. Therefore, type $\theta^*$'s continuation payoff after his first deviation is at most:
\begin{equation}\label{B.11}
    (1-\delta^{T_1})+\delta^{T_1} \Big\{
    (1-\delta^{\overline{k}-1}) +\delta^{\overline{k}} (1-\delta^{\overline{k}-1}) +\delta^{2\overline{k}} (1-\delta^{\overline{k}-1})+...
    \Big\},
\end{equation}
which converges to $\frac{\overline{k}}{1+\overline{k}}$ as $\delta \rightarrow 1$.
Let $k^* \equiv 2\overline{k}$.
When $\delta \rightarrow 1$, type $\theta^*$'s payoff at any on-path history converges to $\frac{2\overline{k}}{2\overline{k}+1}$, which is strictly greater than $\frac{\overline{k}}{1+\overline{k}}$. This verifies type $\theta^*$'s incentive to play his equilibrium strategy.

\section{Concluding Remarks}\label{sec6}
I discuss extensions to environments with imperfect monitoring and non-stationary commitment types.

\paragraph{Imperfect Monitoring:}
Suppose player $2$s observe a noisy signal $y $ distributed according to $f(\cdot| a_{1})$ instead of directly observing $a_{1}$, one can use my proof techniques to establish a weaker lower bound on player $1$'s equilibrium payoff.
Let $\lambda_t \equiv (\lambda_{\theta,t})_{\theta \in \Theta}$ be the likelihood ratio vector with respect to $\alpha_1^* \in \mathcal{A}_1^*$ in period $t$. Let
\begin{equation}\label{6.1}
    \chi_t \equiv \sum_{\theta \in \Theta^b_{(\alpha_1^*,\theta^*)}} \frac{\lambda_{\theta,t}}{\psi^*_{\theta}}.
\end{equation}
When the signal $y$ can statistically identify player $1$'s action $a_{1}$,
type $\theta^*$ player $1$ can secure payoff
\begin{equation}\label{6.2}
    (1-  \chi_0) v_{\theta^*}(\alpha_1^*,u_1,u_2) + \chi_0 \min_{a_2 \in A_2} u_1(\theta^*,\alpha_1^*,a_2)
\end{equation}
when he is sufficiently patient. 
This lower bound is meaningful only when $\chi_0 <1$, or equivalently, when $\lambda (\mu,\alpha_1^*) \in \underline{\Lambda}(\theta^*,\alpha_1^*,u_2)$, in which case player $1$ can secure a fraction $1-\chi_0$ of his commitment payoff. It also incorporates the finding in Fudenberg and Levine (1992), that when player $2$'s best reply to $\alpha_1^*$ does not depend on $\theta$, or equivalently, $\chi_0=0$, player $1$ can secure his commitment payoff from $\alpha_1^*$.

The proof is similar to Statement 3 of Theorem \ref{Theorem1} except that player $1$ cannot perfectly control player $2$s' posterior beliefs due to imperfect monitoring. Nevertheless, $\chi_t$ remains a non-negative supermartingale conditional on the probability measure induced by commitment type $\alpha_1^*$. According to the Doob's Upcrossing Inequality,
the probability of the event that $\chi_t < 1$ for all $t$ is at least $1-\chi_0$. Therefore, type $\theta^*$ can secure at least a fraction $1-\chi_0$ of his commitment payoff from $\alpha_1^*$, regardless of his stage-game payoff function.

\paragraph{Non-Stationary Commitment Types:} In my baseline model, conditional on the state $\theta \in \Theta$ and the commitment plan $\gamma \in \Gamma$, the committed long-run player plays the same action in every period. Thanks to this stationarity assumption, my characterization result for attaining commitment payoff from $\alpha_1^* \in \mathcal{A}_1^*$ does not depend on commitment types playing other actions.

This is not the case when there exist commitment types that play \textit{non-stationary} strategies. For example, suppose there exists a commitment type that mixes between $a_1'$ and $a_1''$ in period $0$, and plays $a_1'$ in all subsequent periods. When examining whether player $1$ can secure his commitment payoff
from $a_1'$, one needs to take into account not only the commitment type that plays $a_1'$ in every period and the state distribution conditional on this commitment type, but also the commitment type that plays the aforementioned nonstationary strategy as well as the state distribution conditional on this non-stationary type. This is because after observing $a_1'$ in period $0$, player $2$s can never distinguish this non-stationary commitment type from commitment type $a_1'$.

One may wonder whether there exists a commitment type (possibly nonstationary), such that player $1$ can secure his optimal commitment payoff
as long as this type occurs with strictly positive probability, regardless of the presence of other commitment types.
The answer to this question is negative as long as player $2$'s best reply to player $1$'s optimal commitment action depends on the state.
This is because the state can be learnt only through the informed player's action choices, not through exogenous signals. For every (potentially non-stationary) commitment plan $\sigma_1^*: \mathcal{H} \times \Theta \rightarrow \Delta (A_1)$, one can construct another commitment plan $\sigma_1^{**}$ that
\begin{enumerate}
  \item Occurs with significantly higher probability compared to $\sigma_1^*$.
  \item Generates the same distribution over public histories as $\sigma_1^*$.
  \item There exists a permutation $\tau: \Theta \rightarrow \Theta$ such that $\sigma_1^* (h^t,\theta) = \sigma_1^{**} (h^t,\tau (\theta))$ for every $(h^t,\theta) \in \mathcal{H} \times \Theta$, that is, the mapping from the states to the committed long-run player's stage-game actions is flipped.
\end{enumerate}
\newpage
\appendix
\section{Proof of Theorem 1: Statement 3}\label{secA}
Throughout this appendix, I suppress the dependence of $a_2^*$, $\lambda$, $\Lambda$, and $\underline{\Lambda}$ on $\alpha_1^*$, $\theta^*$, and $u_2$. Furthermore, I index the set of states by $\{1,2,...,m\}$ instead of $\theta \in \Theta$ when doing summation.
The proof after establishing Corollary 1 consists of two steps.

\subsection{Step 1}\label{subA.3}
For every $\xi>0$, a likelihood ratio vector $\lambda$ is of `\textbf{size} $\boldsymbol{\xi}$' if there exists $\widetilde{\psi} \equiv (\widetilde{\psi}_1,...,\widetilde{\psi}_m) \in \mathbb{R}_+^m$ such that: $\widetilde{\psi}_i \in (0,\psi_i)$ for all $i$ and moreover,
\begin{equation}\label{B.4}
    \lambda \in \Big\{\widetilde{\lambda} \in \mathbb{R}_+^m \Big|
    \sum_{i=1}^m \widetilde{\lambda}_i/\widetilde{\psi}_i <1
    \Big\} \subset \Big\{\widetilde{\lambda} \in \mathbb{R}_+^m \Big|
    \# \{i|\widetilde{\lambda}_i \leq \xi\} \geq m-1
    \Big\}.
\end{equation}
Intuitively, $\lambda$ is of size $\xi$ if there exists a downward sloping hyperplane such that every non-negative likelihood ratio vector below this hyperplane has at least $m-1$ entries no larger than $\xi$. By definition, for every $\xi' \in (0,\xi)$, if $\lambda$ is of size $\xi'$, then it is also of size $\xi$. Proposition \ref{Prop2} establishes (\ref{3.3}) when $\lambda$ is of size $\xi$ for $\xi$ small enough.
\begin{Proposition}\label{Prop2}
There exists $\xi>0$, s.t.
$\lim \inf_{\delta \rightarrow 1} \underline{v}_{\theta^*} (\delta,\mu,u_1,u_2) \geq u_1(\theta^*,\alpha_1^*,a_2^*)$ for every   $\lambda$  of size $\xi$.
\end{Proposition}
\noindent\textsc{Proof:}~
Let $\alpha_1(\cdot|h^t,\omega_i)\in \Delta(A_1)$ be the equilibrium action of type $\omega_i$ at history $h^t$. Let
\begin{equation}\label{B.5}
    B_{i,a_1} (h^t) \equiv \lambda_i(h^t) \Big(\alpha_1^*(a_1)-\alpha_1(a_1|h^t,\omega_i)\Big).
\end{equation}
Recall that $\alpha_1(\cdot|h^t)$
is the average action expected by player $2$.
For every $\lambda \in \underline{\Lambda}(\alpha_1^*,\theta^*,u_2)$ and $\epsilon>0$, there exists $\varepsilon>0$ such that
for every likelihood ratio vector $\widetilde{\lambda}$ satisfying:
\begin{equation}\label{B.6}
    \sum_{i=1}^m \widetilde{\lambda}_i/\psi_i <\frac{1}{2}\Big(1+\sum_{i=1}^m \lambda_i/\psi_i\Big),
\end{equation}
$a_2^*$ is player $2$'s strict best reply to every
$\{\alpha_1(\cdot|h^t,\omega_i)\}_{i=1}^m$ satisfying the following two conditions
\begin{itemize}
\item[1.] $|B_{i,a_1} (h^t)|<\varepsilon$ for all $i$ and $a_1$.
\item[2.] $\big\| \alpha^*_1 - \alpha_1(\cdot|h^t) \big\| \leq \epsilon$.
\end{itemize}
This is because when the prior likelihood ratio vector satisfies (\ref{B.6}), $a_2^*$  is player $2$'s strict best reply when all types of player $1$ play $\alpha_1^*$. When $\epsilon$ and $\varepsilon$ are both small enough, an $\epsilon$-deviation of the average action together with an $\varepsilon$-correlation between types and actions cannot overturn this strictness.

According to the Pinsker's Inequality, $\big\| \alpha^*_1 - \alpha_1(\cdot|h^t) \big\| \leq \epsilon$ is implied by $d(\alpha^*_1||\alpha_1(\cdot|h^t)) \leq \epsilon^2/2$. Pick $\epsilon$ and $\xi$ small enough such that:
\begin{equation}\label{B.7}
    \epsilon <\frac{\varepsilon}{2(1+\overline{\psi})} \quad
\textrm{and} \quad
    \xi < \frac{\varepsilon}{(m-1)(1+\varepsilon)}.
\end{equation}
Suppose $\lambda_i(h^t) \leq \xi$ for all $i \geq 2$, since $\big\| \alpha^*_1 - \alpha_1(\cdot|h^t) \big\| \leq \epsilon$, we have:
\begin{equation*}
    \frac{\displaystyle \Big\|
  \lambda_1 (\alpha_1^*-\alpha_1(a_1|h^t,\omega_1))+\sum_{i=2}^m \lambda_i \big(\alpha_1^*-\alpha_1(a_1|h^t,\omega_i)\big)
    \Big\|}{1+\lambda_1+\xi (m-1)} \leq \epsilon.
\end{equation*}
The triangular inequality implies that:
\begin{eqnarray}\label{B.9}
    \Big\|
  \lambda_1 (\alpha_1^*-\alpha_1(a_1|h^t,\omega_1)) \Big\|
 & \leq & \sum_{i=2}^m \Big\|
  \lambda_i (\alpha_1^*-\alpha_1(a_1|h^t,\omega_i)) \Big\|+\epsilon \Big(
  1+\lambda_1+\xi (m-1)
  \Big)
        {}
\nonumber\\
& \leq & \xi (m-1)+\epsilon  \Big(
  1+\overline{\psi}+\xi (m-1)
  \Big) \leq \varepsilon.
\end{eqnarray}
where the last inequality uses (\ref{B.7}). Inequality (\ref{B.9}) implies that
$||B_{1,a_1}(h^t)|| \leq \varepsilon$.
As a result,
for every $\lambda$ of size $\xi$, $a_2^*$ is player $2$'s strict best reply at every history $h^t$ satisfying $d(\alpha_1^*||\alpha_1(\cdot|h^t)) \leq \epsilon^2/2$.
\qed

\subsection{Step 2}\label{subA.4}
I apply the conclusion of Proposition \ref{Prop2} to establish inequality (\ref{3.3}) for every $\lambda \in \underline{\Lambda}$.
Recall the definition of $B_{i,a_1} (h^t)$ in (\ref{B.5}). According to Bayes rule, if player $1$ plays $a_1 \in A_1^*$ at $h^t$, then
\begin{equation*}
   \lambda_i(h^t)-\lambda_i(h^t,a_1)=\frac{B_{i,a_1} (h^t)}{\alpha_1^*(a_1)} \quad
\textrm{and }
    \sum_{a_1 \in A_1^*} \alpha_1^*(a_1)  \Big( \lambda_i(h^t)-\lambda_i(h^t,a_1) \Big)\geq 0.
\end{equation*}
Let
$D(h^t,a_1) \equiv  \Big(\lambda_i(h^t)-\lambda_i(h^t,a_1) \Big)_{i=1}^m \in \mathbb{R}^m$.
Suppose $B_{i,a_1} (h^t) \geq \varepsilon$ for some $i$ and $a_1 \in A_1^*$, we have $||D(h^t,a_1)|| \geq \varepsilon$
 where $||\cdot||$ denotes the $\mathcal{L}^2$-norm.
Pick $\xi>0$ small enough to meet the requirement in Proposition \ref{Prop2}.
I define two sequences of subsets of $\underline{\Lambda}(\theta^*,\alpha_1^*,u_2)$, namely  $\{\Lambda^k\}_{k=0}^{\infty}$ and $\{\widehat{\Lambda}^k\}_{k=1}^{\infty}$:
\begin{itemize}
  \item Let $\Lambda^0$ be the set of likelihood ratio vectors that are of size $\xi$,
  \item For every $k \geq 1$, let $\widehat{\Lambda}^k$ be the set of likelihood ratio vectors in $\underline{\Lambda}(\theta^*,\alpha_1^*,u_2)$ such that if $\lambda(h^t) \in \widehat{\Lambda}^k$, then either $\lambda(h^t) \in \Lambda^{k-1}$ or, For every $\{\alpha_1(\cdot|h^t,\omega_i)\}_{i=1}^m$ such that $||D(h^t,a_1)|| \geq \varepsilon$ for some  $a_1 \in A_1^*$, there exists $a_1^* \in A_1^*$ such that $\lambda(h^t,a_1^*) \in \Lambda^{k-1}$.
\item Let $\Lambda^k$ be the set of likelihood ratio vectors in $\underline{\Lambda}(\theta^*,\alpha_1^*,u_2)$ such that  for every $\tilde{\lambda} \in \Lambda^k$, there exists $\tilde{\psi} \equiv (\tilde{\psi}_1,...,\tilde{\psi}_m) \in \mathbb{R}_+^m$ such that: $\tilde{\psi}_i \in (0,\psi_i)$ for all $i$ and
\begin{equation}\label{B.10}
    \lambda \in \Big\{\tilde{\lambda} \in \mathbb{R}_+^m \Big|
    \sum_{i=1}^m \tilde{\lambda}_i/\tilde{\psi}_i <1
    \Big\} \subset \Big(\bigcup_{j=0}^{k-1} \Lambda^j \Big) \bigcup \widehat{\Lambda}^k.
\end{equation}
By construction,
\begin{equation}\label{B.11}
    \Big\{\tilde{\lambda} \in \mathbb{R}_+^m \Big|
    \sum_{i=1}^m \tilde{\lambda}_i/\tilde{\psi}_i <1
    \Big\} \subset \bigcup_{j=0}^{k} \Lambda^j =\Lambda^k.
\end{equation}
\end{itemize}
Since $(0,...,\psi_i-\upsilon,...,0) \in \Lambda^0$ for any $i \in \{1,2,...,m\}$ and $\upsilon>0$, so $\textrm{co} (\Lambda^0)= \underline{\Lambda}(\theta^*,\alpha_1^*,u_2)$. By definition, $\{\Lambda^k\}_{k \in \mathbb{N}}$ is an increasing sequence with $\Lambda^k \subset \underline{\Lambda}(\theta^*,\alpha_1^*,u_2) =\textrm{co}(\Lambda^k)$ for any $k \in \mathbb{N}$, i.e. it is bounded from above by a compact set. Therefore $\lim_{k \rightarrow \infty}  \bigcup_{j=0}^{k} \Lambda^j \equiv \Lambda^{\infty}$ exists and is a subset of $\textrm{cl}\Big(\underline{\Lambda}(\theta^*,\alpha_1^*,u_2) \Big)$. The next Lemma shows that  $\textrm{cl}(\Lambda^{\infty})$ coincides with $\textrm{cl}\Big(\underline{\Lambda}(\theta^*,\alpha_1^*,u_2) \Big)$.
\begin{Lemma}\label{L2}
$\textrm{cl}(\Lambda^{\infty})=\textrm{cl}\Big(\underline{\Lambda}(\theta^*,\alpha_1^*,u_2) \Big)$
\end{Lemma}
\noindent\textsc{Proof of Lemma A.1:}~
Since $\Lambda^k \subset \underline{\Lambda}(\theta^*,\alpha_1^*,u_2)$ for every $k \in \mathbb{N}$, $\textrm{cl}(\Lambda^{\infty})\subset \textrm{cl}\Big(\underline{\Lambda}(\theta^*,\alpha_1^*,u_2)\Big)$. The rest of the proof shows the other direction. Suppose by way of contradiction that
    $\textrm{cl}(\Lambda^{\infty}) \subsetneq \textrm{cl}\Big(\underline{\Lambda}(\theta^*,\alpha_1^*,u_2) \Big)$.
\begin{itemize}
\item[1.] Let $\widehat{\Lambda}\subset \underline{\Lambda}(\theta^*,\alpha_1^*,u_2)$ be such that if $\lambda(h^t) \in \widehat{\Lambda}$, then \textit{either} $\lambda(h^t) \in \Lambda^{\infty}$, \textit{or}
 for every $\{\alpha_1(\cdot|h^t,\omega_i)\}_{i=1}^m$ such that $||D(h^t,a_1)|| \geq \varepsilon$ for some  $a_1 \in A_1^*$, there exists $a_1^* \in A_1^*$ such that $\lambda(h^t,a_1^*) \in \Lambda^{\infty}$.
\item[2.] Let $\breve{\Lambda}$ be the set of likelihood ratio vectors in $\underline{\Lambda}(\theta^*,\alpha_1^*,u_2)$ such that  for every $\widetilde{\lambda} \in \breve{\Lambda}$, there exists $\widetilde{\psi} \equiv (\widetilde{\psi}_1,...,\widetilde{\psi}_m) \in \mathbb{R}_+^m$ such that:
\begin{equation}\label{B.13}
\widetilde{\psi}_i \in (0,\psi_i) \textrm{ for all $i$ and }    \lambda \in \Big\{\widetilde{\lambda} \in \mathbb{R}_+^m \Big|
    \sum_{i=1}^m \widetilde{\lambda}_i/\widetilde{\psi}_i <1
    \Big\} \subset  \Big(\Lambda^{\infty} \bigcup \widehat{\Lambda} \Big).
\end{equation}
\end{itemize}
Since $\Lambda^{\infty}$ is defined as the limit of the above operator, so in order for $\textrm{cl}(\Lambda^{\infty}) \subsetneq \textrm{cl}\Big(\underline{\Lambda}(\theta^*,\alpha_1^*,u_2) \Big)$ to be true, it has to be the case that $\breve{\Lambda}=\Lambda^{\infty}$, or $\Xi \bigcap \breve{\Lambda}=\{\varnothing\}$ where
\begin{equation}\label{B.14}
    \Xi \equiv \textrm{cl}\Big(\underline{\Lambda}(\theta^*,\alpha_1^*,u_2) \Big) \Big\backslash  \textrm{cl}(\Lambda^{\infty}).
\end{equation}
One can check that $\Xi$ is convex and has non-empty interior. For every $\varrho>0$, there exists $x \in \Xi$, $\theta \in (0,\pi/2)$ and a halfspace
$H(\chi) \equiv \Big\{\widetilde{\lambda} \Big|   \sum_{i=1}^m \widetilde{\lambda}_i/\chi_i < \chi \Big\}$
with $\phi > 0$ satisfying:
\begin{enumerate}
  \item $\sum_{i=1}^m x_i/\psi_i=\chi$.
  \item $\partial B(x,r) \bigcap H(\chi) \bigcap \underline{\Lambda}(\theta^*,\alpha_1^*,u_2) \subset \Lambda^{\infty}$ for every $r \geq \varrho$.
  \item For every $r \geq \rho$ and $y \in \partial B(x,r) \bigcap \underline{\Lambda}(\theta^*,\alpha_1^*,u_2)$, either $y \in \Lambda^{\infty}$ or $d(y,H(\chi))>r \sin \theta$, where $d(\cdot,\cdot)$ denotes the Hausdorff distance.
\end{enumerate}
The second and third property used the non-convexity of $\textrm{cl}(\Lambda^{\infty})$.
Suppose $\lambda(h^t)=x$ for some $h^t$ and there exists $a_1 \in A_1^*$ such that $||D(h^t,a_1)|| \geq \varepsilon$,
\begin{itemize}
  \item Either $\lambda(h^t,a_1) \in \Lambda^{\infty}$, in which case $x \in \breve{\Lambda}$ but $x \in \Xi$, leading to a contradiction.
  \item Or $\lambda(h^t,a_1) \notin \Lambda^{\infty}$. Requirement 3 implies that $d(\lambda(h^t,a_1),H(\chi))> \varepsilon \sin \theta$. On the other hand,
      \begin{equation}\label{B.15}
        \sum_{a_1' \in A_1^*} \alpha_1^*(a_1') \lambda_i(h^t,a_1') \leq \lambda_i(h^t)
      \end{equation}
      for every $i$. Requirement 1 then implies that $\sum_{a_1' \in A_1^*} \alpha_1^*(a_1')
         \lambda_i(h^t,a_1') \in H(\chi)$,
      which is to say:
      \begin{equation}\label{B.16}
        \sum_{a_1' \in A_1^*} \alpha_1^*(a_1')
        \sum_{i=1}^m
         \lambda_i(h^t,a_1')/\psi_i \leq \chi.
      \end{equation}
According to Requirement 2, $\lambda(h^t,a_1) \notin H(\chi)$. In another word,
$\sum_{i=1}^m \lambda_i(h^t,a_1)/\psi_i > \chi+\varepsilon \kappa$ for some $\kappa >0$. Let
    $\rho \equiv\frac{1}{2} \min_{a_1 \in A_1^*} \{\alpha_1^*(a_1)\}  \varepsilon \kappa$.
(\ref{B.15}) implies the existence of $a_1^* \in A_1^* \backslash \{a_1\}$ such that $\lambda(h^t,a_1^*) \in H(\chi) \bigcap B(x,\rho)$. Requirement 2 then implies that $x=\lambda(h^t) \in \breve{\Lambda}$. Since $x \in \Xi$, this leads to a contradiction and validates the conclusion of Lemma \ref{L2}.
\end{itemize}
\qed

Lemma \ref{L2} implies that for every $\lambda \in \underline{\Lambda}(\theta^*,\alpha_1^*,u_2)$, there exists an integer $K \in \mathbb{N}$ independent of $\delta$ such that $\lambda \in \Lambda^K$.
Statement 3 of Theorem 1 can then be shown by induction on $K$. According to Proposition \ref{Prop2}, the statement holds for $K=0$. Suppose it applies to every $K \leq K^*-1$, let us consider the case when $K=K^*$. According to the construction of $\Lambda^{K^*}$, there exists a strategy for player $1$ such that whenever $a_2^*$ is not player $2$'s best reply despite $d(\alpha_1^*\|\alpha_1(\cdot|h^t))<\epsilon^2/2$, then the posterior belief after observing $a_{1,t}$ is in $\Lambda^{K^*-1}$, under which the commitment payoff bound is attained by the induction hypothesis.
\section{Proof of Lemma 4.1}\label{secB}
For every $n \in \mathbb{N}$,
let $\widehat{X}_n \equiv \delta^n (X_n-\alpha_1^*(a_1))$. Define a triangular sequence of random variables $\{X_{k,n}\}_{0 \leq n\leq k, k,n \in \mathbb{N}}$, such that $X_{k,n}\equiv \xi_k \widehat{X}_n$, where
    $\xi_k \equiv \sqrt{\frac{1}{\sigma^2} \frac{1-\delta^2}{1-\delta^{2k}}}$.
Let $Z_k \equiv \sum_{n=1}^k X_{k,n} =\xi_k \sum_{k=1}^n \widehat{X}_n$.
According to the Lindeberg-Feller Central Limit Theorem, $Z_k$ converges in law to $ N(0,1)$. By construction,
\begin{equation*}
    \frac{\sum_{n=1}^k \widehat{X}_n}{1+\delta+...+\delta^{k-1}}=\sigma   \sqrt{\frac{1-\delta^{2k}}{1-\delta^2}} \frac{1-\delta}{1-\delta^k} Z_k.
\end{equation*}
The RHS of this expression converges in law to a normal distribution with mean $0$ and variance
$\sigma^2 \frac{1-\delta^{2k}}{1-\delta^2} \frac{(1-\delta)^2}{(1-\delta^k)^2}$.
The variance term converges to $\mathcal{O}\Big((1-\delta)\Big)$ as $k \rightarrow \infty$. According to Theorem 7.4.1 in Chung (1974), we have:
\begin{equation*}
    \sup_{x \in \mathbb{R}} |F_k(x)-\Phi(x)| \leq C_0 \sum_{n=1}^{k} |X_{k,n}|^3 \sim C_1 (1-\delta)^{\frac{3}{2}},
\end{equation*}
where $C_0$ and $C_1$ are constants, $F_k$ is the empirical distribution of $Z_k$ and $\Phi(\cdot)$ is the cdf of the standard normal distribution. Both the variance and the approximation error converge to $0$ as $\delta \rightarrow 1$.

Therefore, for every $\eta>0$, there exists $\overline{\delta} \in (0,1)$ such that for every $\delta >\overline{\delta}$, there exists $K \in \mathbb{N}$, such that for all $k>K$,
\begin{equation*}
    \mathcal{P}^{(\alpha_1^*,\sigma_2)} \Big(
   \Big| \frac{\sum_{i=1}^k \widehat{X}_n}{1+\delta+...+\delta^{k-1}} \Big| \geq \eta
    \Big)<  \frac{\eta}{n}.
\end{equation*}
The conclusion of Lemma \ref{L4.2} is obtained by taking $k \rightarrow \infty$.

\end{spacing}

\end{document}